\documentclass[
]{ceurart}

\usepackage[english]{babel}
\usepackage{soul}
\usepackage{adjustbox}
\usepackage{url}
\usepackage{graphicx}
\usepackage{booktabs}
\usepackage{pifont}
\usepackage{listings}
\usepackage{todonotes}
\usepackage{tikz}
\usetikzlibrary{arrows,arrows.meta,calc}
\usetikzlibrary{shadows}
\usetikzlibrary{shapes.multipart, shapes.geometric}
\usetikzlibrary{positioning}
\usetikzlibrary{shadows}
\usetikzlibrary{fit}
\usetikzlibrary{shapes.multipart}
\usetikzlibrary{arrows.meta}
\usetikzlibrary{graphs,quotes}
\usepackage{xcolor-solarized}

\usepackage{cleveref}

\usepackage[normalem]{ulem}
\usepackage{xcolor}
\definecolor{myyellow}{RGB}{255,245,170}
\sethlcolor{myyellow}

\begin{document}

\copyrightyear{2026}
\copyrightclause{Copyright for this paper by its authors.
  Use permitted under Creative Commons License Attribution 4.0
  International (CC BY 4.0).}

\conference{2nd International Workshop on Data Management for Knowledge Graphs (DMKG 2026)}

\title{sheval: An RDF data shapes evaluation tool and test-suite for recursive shapes}


\author[1]{Shqiponja Ahmetaj}
\author[2]{Iovka Boneva}
\author[3]{Jan Hidders}
\author[1]{Maxime Jakubowski}
\author[4]{Jose-Emilio Labra-Gayo}[%
orcid=0000-0001-8907-5348,
url=https://labra.weso.es/,
]
\cormark[1]
\author[5]{Wim Martens}
\author[6]{Filip Murlak}
\author[7]{Cem Okulmus}[%
orcid=0000-0002-7742-0439,
]
\author[8]{Ognjen Savkovi\'c}
\author[1]{Mantas \v{S}imkus}
\author[9]{Dominik Tomaszuk}[%
orcid=0000-0003-1806-067X,
]

\address[1]{TU Wien, Austria}
\address[2]{Univ. Lille, F-59000 Lille, France}
\address[3]{Birkbeck, University of London, UK}
\address[4]{University of Oviedo, Spain}
\address[5]{University of Bayreuth, Germany}
\address[6]{University of Warsaw, Poland}
\address[7]{Paderborn University, Germany}
\address[8]{Free University of Bolzano, Italy}
\address[9]{University of Białystok, Poland}

\cortext[1]{Corresponding author.}

\begin{abstract}
  Two different languages have been developed to validate RDF data based on the concept of a shape: ShEx and SHACL.
  In each language it is possible to define a shape that refers to itself, which is called a recursive shape.
  While in the case of ShEx, the semantics of recursive shapes is well defined and is part of the specification, in the case of SHACL, 
   the semantics of recursive shapes is left to the implementation of the different SHACL engines.
  Consequently, the different SHACL engines show different behaviours when confronted with recursive shapes.
  In this paper we present sheval: an evaluation framework consisting of a tool and a test suite 
   that can be used to compare the behaviour of different shapes technologies 
  when confronted with recursive definitions.
  The tool has been used to evaluate and understand the differences in the implementation of recursive shapes in ShEx and SHACL.
  It provides a framework for testing and comparing the behaviour of different shape engines, 
  helping to identify inconsistencies and potential issues, and providing a basis for further research and development 
  in the field of shape-based validation of RDF data.
\end{abstract}



\newtheorem{definition}{Definition}
\newtheorem{construction}{Construction}
\newtheorem{proposition}{Proposition}
\newtheorem{lemma}{Lemma}
\newtheorem{claim}{Claim}
\newtheorem{theorem}{Theorem}
\newtheorem{corollary}{Corollary}
\newtheorem{conjecture}{Conjecture}
\newtheorem{remark}{Remark}
\newtheorem{example}{Example}


\newcommand{\mcomment}[2]{}
\newcommand{\footcomment}[2]{{\color{blue}\textbf{(#1)}}\footnote{\textbf{#1:} #2}}
\newcommand{\margincomment}[2]{{\color{blue}\textbf{(#1)}}\footnotemark\marginnote{\tiny\textsuperscript{\thefootnote}\textbf{#1:} #2}}


\newcommand{\wim}[1]{\mcomment{Wim}{#1}}
\newcommand{\jose}[1]{\mcomment{Jose}{#1}}
\newcommand{\maxime}[1]{\mcomment{Maxime}{#1}}
\newcommand{\filip}[1]{\mcomment{Filip}{#1}}
\newcommand{\mantas}[1]{\mcomment{Mantas}{#1}}
\newcommand{\fabio}[1]{\mcomment{Fabio}{#1}}
\newcommand{\iovka}[1]{\mcomment{Iovka}{#1}}
\newcommand{\jan}[1]{\mcomment{Jan}{#1}}
\newcommand{\cem}[1]{\mcomment{Cem}{#1}}
\newcommand{\dominik}[1]{\mcomment{Dominik}{#1}}
\newcommand{\ognjen}[1]{\mcomment{Ognjen}{#1}}
\newcommand{\shqiponja}[1]{\mcomment{Shqiponja}{#1}}



\newcommand{\alogspace}{\mathsc{ALogSpace}}
\newcommand{\ptime}{\mathsc{P}}
\newcommand{\nptime}{\mathsc{NP}}

\newcommand{\graph}{G}
\newcommand{\schema}{\mathcal{S}}
\newcommand{\cat}{C}
\newcommand{\decl}[2]{#1 \colon\! #2}
\newcommand{\Domain}{\Delta}
\newcommand{\sel}{\mathit{sel}}


\newcommand{\Nodes}{\nodes}
\newcommand{\Values}{\mathcal{V}}
\newcommand{\Labels}{\mathcal{L}}
\newcommand{\Predicates}{\mathcal{P}}
\newcommand{\Keys}{\mathcal{K}}
\newcommand{\Records}{\mathcal{R}}
\newcommand{\nodes}{\mathsf{Nodes}}
\newcommand{\edges}{\mathsf{Edges}}
\newcommand{\predicates}{\mathsf{Predicates}}
\newcommand{\keys}{\mathsf{Keys}}
\newcommand{\values}{\mathsf{Values}}

\newcommand{\Names}{\mathsf{Names}}

\newcommand{\IRIs}{\mathsf{IRIs}}
\newcommand{\Blanks}{\mathsf{Blanks}}
\newcommand{\Literals}{\mathsf{Literals}}

\newcommand{\RDFtype}{\mathsf{type}}




\newcommand{\mathsc}[1]{{\normalfont\textsc{#1}}}

\newcommand{\Exkey}[1]{\mathit{#1}}
\newcommand{\Exprop}[1]{\mathsf{#1}}

\newcommand{\exowns}{\Exprop{ownsAccount}}
\newcommand{\exaccess}{\Exprop{hasAcccess}}
\newcommand{\exinvited}{\Exprop{invited}}
\newcommand{\exemail}{\Exkey{email}}
\newcommand{\excard}{\Exkey{card}}
\newcommand{\exprivileged}{\Exkey{privileged}}

\newcommand{\key}{SHACL shape\xspace}

\newcommand{\OMIT}[1]{}

\newcommand{\shapeTerm}{SHACL shape\xspace}
\newcommand{\selTerm}{SHACL selector\xspace}
\newcommand{\SHACLSchema}{\schema}
\newcommand{\pathExpr}{\pi}
\newcommand{\SHACLShapeDef}{\mathit{D}}
\newcommand{\SHACLShapeMap}{\mathit{M}}

\newcommand{\id}{\mathsf{id}}
\newcommand{\eq}{\mathsf{eq}}
\newcommand{\disj}{\mathsf{disj}}
\newcommand{\geqn}[2]{\exists^{\geq #1}#2.}
\newcommand{\leqn}[2]{\exists^{\leq #1}#2.}
\newcommand{\hasshape}{\mathsf{hasShape}}
\newcommand{\hasvalue}{\mathsf{cond}}
\newcommand{\test}{\mathsf{test}}
\newcommand{\closed}{\mathsf{closed}}
\newcommand{\lessthan}{\mathsf{lessThan}}
\newcommand{\lessthaneq}{\mathsf{lessThanEq}}
\newcommand{\morethan}{\mathsf{moreThan}}
\newcommand{\morethaneq}{\mathsf{moreThanEq}}
\newcommand{\eqlang}{\sim}
\newcommand{\uniquelang}{\mathit{uniqueLang}}
\newcommand{\iexpr}[2]{\llbracket #1 \rrbracket^{#2}}
\newcommand{\enne}{\ \& \ }
\newcommand{\typesymb}{\ell}



\newcommand{\PGSchema}{\mathcal{S}_{\text{pg}}}
\newcommand{\PGTypeNameSet}{\textit{TN}}
\newcommand{\NodeTypeSet}{S}
\newcommand{\EdgeTypeSet}{T}
\newcommand{\PGConstraintSet}{C}

\newcommand{\sem}[1]{\llbracket{#1}\rrbracket}
\newcommand{\semdelta}[3]{\llbracket{#1}\rrbracket^{#2}_{#3,\cat}}
\newcommand{\gsem}[1]{\llbracket{#1}\rrbracket^{\graph}}
\newcommand{\esem}[1]{\llbracket{#1}\rrbracket^{\graph}_{\textsf{el}}}
\newcommand{\vsem}[1]{\llbracket{#1}\rrbracket_{\textsf{val}}}

\newcommand{\opensym}{\circ}  
\newcommand{\closedsym}{\bullet}   

\newcommand{\rp}[1]{\texttt{\{} #1 \texttt{\}}}  
\newcommand{\openRT}[1]{\rp{#1}^{\circ}}  
\newcommand{\closedRT}[1]{\rp{#1}}  
\newcommand{\emptyRec}{\textbf{r}_{\emptyset}}

\newcommand{\tOr}{\mathbin{\texttt{|}}}
\newcommand{\tAnd}{\mathbin{\texttt{\&}}}

\newcommand{\pgschemas}{\textsc{PG-Schemas}\xspace}
\newcommand{\pgschema}{\textsc{PG-Schema}\xspace}
\newcommand{\pgtypes}{\textsc{PG-Types}\xspace}
\newcommand{\pgtype}{\textsc{PG-Type}\xspace}
\newcommand{\pgkeys}{\textsc{PG-Keys}\xspace}
\newcommand{\pgkey}{\textsc{PG-Key}\xspace}

\newcommand{\cgschemas}{\textsc{CG-Schemas}\xspace}
\newcommand{\cgschema}{\textsc{CG-Schema}\xspace}
\newcommand{\cgtypes}{\textsc{CG-Types}\xspace}
\newcommand{\cgtype}{\textsc{CG-Type}\xspace}
\newcommand{\cgkeys}{\textsc{CG-Keys}\xspace}
\newcommand{\cgkey}{\textsc{CG-Key}\xspace}

\newcommand{\ValueTypes}{\mathcal{T}} 
\newcommand{\vtype}{\tau} 

\newcommand{\NodeTypes}{\mathcal{T}_{\textbf{n}}}  

\newcommand{\gDef}{{\color{orange} \ ::= \ }}
\newcommand{\gMid}{{\color{orange} \ \big|\ }}
\newcommand{\gEnd}{{\color{orange} \ .\ }}
\newcommand{\gOpt}{{\color{orange}?}}

\newcommand{\pexpr}{\pi}
\newcommand{\ppexpr}{\bar{\pi}}

\newcommand{\content}[2]{\textbf{K}_{#2}({#1})}  
\newcommand{\et}[3]{{#1}\stackrel{#2}{\rightarrow}{#3}}  
\newcommand{\pwc}{\star}  
\newcommand{\lab}[1]{\texttt{#1}}  
\newcommand{\PredTypes}{\mathcal{T}_{\textsf{p}}}
\newcommand{\EdgeTypes}{\mathcal{T}_{\textsf{e}}}

\newcommand{\Key}{\textsf{\bf Key}}
\newcommand{\mand}{\textsf{\bf mnd}}
\newcommand{\sing}{\textsf{\bf sng}}

\newcommand{\keyIsVal}[2]{[{#1}={#2}]}

\newcommand{\shexsel}{\mathit{sel}}
\newcommand{\shexneigh}[1]{\big\{ #1 \big\}}
\newcommand{\shexneighzero}[1]{\shexneigh{#1 \shexeach \top}}
\newcommand{\shexneighopen}[1]{\left\{ #1 \right\}^{\circ}}

\newcommand{\shexref}{}
\newcommand{\shexeach}{\mathop{;}}
\newcommand{\shexone}{\mathop{|}}
\newcommand{\shexinverse}[1]{#1^{-}}
\newcommand{\shexneg}[1]{\neg #1}
\newcommand{\shexneginv}[1]{\neg{\shexinverse{#1}}}

\newcommand{\ttopen}{\textit{op}_{\pm}}
\newcommand{\ttclosed}{\textit{op}_{-}}

\newcommand{\mleft}{\mathopen{\{\hspace{-0.25em}|}}
\newcommand{\mright}{\mathclose{|\hspace{-0.25em}\}}}
\newcommand{\msetin}{\mathbin{\in\hspace{-0.45em}\in}}
\newcommand{\proj}{\mathrel{\rightsquigarrow}}
\newcommand{\projlang}{\mathit{Proj^{-1}}}
\newcommand{\FF}{\mathcal{F}}
\newcommand{\Shexedgelabels}{\mathcal{Q}}
\newcommand{\shexmultisets}[1]{\mathbb{N}^{#1}}

\newcommand{\neigh}{\mathsf{Neigh}}



\newcommand{\Triples}{\mathcal{E}}

\newcommand{\const}{\mathsf{Const}}

\newcommand{\stable}{\textsf{st}}
\newcommand{\supp}{\textsf{supp}}

\newcommand{\SHACLpath}[1]{\text{SHACL}^{\pi}_{#1}}
\newcommand{\SHACL}[1]{\text{SHACL}_{#1}}
\newcommand{\mSHACL}{\text{mSHACL}}

\newcommand{\SHEX}[1]{\text{ShEx}_{#1}}
\newcommand{\mSHEX}{\text{mShEx}}



\newcommand{\microshex}{\ensuremath{\text{ShEx}_0}}
\newcommand{\microshacl}{\ensuremath{\text{SHACL}_0}}

\newcommand{\LFP}{LFP}
\newcommand{\GFP}{GFP}
\newcommand{\SMS}{SMS}

\newcommand{\transl}{\tau}
\newcommand{\illustration}[1]{}
\newcommand{\removableforspace}[1]{\textcolor{black!50}{#1}}

\newcommand{\testcode}[1]{\texttt{#1}}
\newcommand{\pass}{\color{OliveGreen}}
\newcommand{\fail}{\color{Brown}}
\newcommand{\error}{\color{BrickRed}}
\newcommand{\stopped}{\color{RoyalBlue}}

\newcommand{\cmark}{\pass \phantom{a}\ding{51}}%
\newcommand{\xmark}{\fail \phantom{a}\ding{55}\phantom{*}}%
\newcommand{\xxmark}{\fail \phantom{a}\ding{55}*}%


\colorlet{RDFComment}{gray}
\colorlet{RDFString}{OliveGreen}
\colorlet{RDFPrefix}{Purple}
\colorlet{RDFKeyword}{MidnightBlue}
\colorlet{RDFIRI}{teal!70!black}


\def\RDFPrefixes{
  rdf,rdfs,owl,xsd,
  sh,shex,
  skos,skosxl,
  schema,
  foaf,
  dc,dct,
  qb,void,
  prov,
  org,
  doap,
  lemon,
  ex
}


\lstdefinestyle{rdfstyle}{
  basicstyle=\ttfamily,
  keepspaces=true,
  columns=flexible,
  breaklines=false,
  sensitive=true,
  frame=bt,
  aboveskip=1em,
  belowskip=1em,
  xleftmargin=.5em,
  xrightmargin=.5em,
  framexleftmargin=.5em,
  framexrightmargin=.5em,
  framextopmargin=.5em,
  framexbottommargin=.5em,
  tabsize=2,
  showstringspaces=false,
  commentstyle=\color{RDFComment},
  stringstyle=\color{RDFString}
}


\lstdefinelanguage{Turtle}{
  style=rdfstyle,
  alsoletter={:,@,-},
  morecomment=[l]{\#},
  morecomment=[n][\color{RDFIRI}]{<}{>},
  morestring=[b]',
  morestring=[b]",
  classoffset=1,
  keywordstyle=\color{RDFPrefix},
  morekeywords={\RDFPrefixes},
  classoffset=2,
  keywordstyle=\color{RDFKeyword},
  morekeywords={
    @prefix,
    @base,
    PREFIX,
    BASE,
    a,
    true,
    false
  },
  classoffset=0
}


\lstdefinelanguage{SHACL}{
  style=rdfstyle,
  alsoletter={:,@,-},
  morecomment=[l]{\#},
  morecomment=[n][\color{RDFIRI}]{<}{>},
  morestring=[b]',
  morestring=[b]",
  classoffset=1,
  keywordstyle=\color{RDFPrefix},
  morekeywords={\RDFPrefixes},
  classoffset=2,
  keywordstyle=\color{RDFKeyword},
  morekeywords={
    @prefix,@base,
    PREFIX,BASE,
    a,
  },
  classoffset=0,
  literate=
    {sh:}{{{\color{RDFKeyword}sh:}}}3
    {rdf:}{{{\color{RDFKeyword}rdf:}}}4
}


\lstdefinelanguage{ShExC}{
  style=rdfstyle,
  alsoletter={:,@,-},
  morecomment=[l]{\#},
  morecomment=[n][\color{RDFIRI}]{<}{>},
  morestring=[b]',
  morestring=[b]",
  classoffset=1,
  keywordstyle=\color{RDFPrefix},
  morekeywords={\RDFPrefixes},
  classoffset=2,
  keywordstyle=\color{RDFKeyword},
  morekeywords={
    PREFIX,
    BASE,
    IMPORT,
    START,
    CLOSED,
    EXTRA,
    ABSTRACT,
    EXTERNAL,
    EXTENDS,
    RESTRICTS,
    AND,
    OR,
    NOT,
    IRI,
    BNODE,
    LITERAL,
    NONLITERAL,
    MININCLUSIVE,
    MAXINCLUSIVE,
    MINEXCLUSIVE,
    MAXEXCLUSIVE
  },
  classoffset=0
}


\lstdefinelanguage{YAML}{
  style=rdfstyle,
  sensitive=true,
  alsoletter={:,-,_},
  morecomment=[l]{\#},
  morestring=[b]',
  morestring=[b]",
  classoffset=2,
  keywordstyle=\color{RDFKeyword},
  morekeywords={
    true,false,
    yes,no,
    on,off,
    null,NULL,Null,
    ~
  },
  literate=
    {:}{{{\color{RDFKeyword}:}}}1
    {-}{{{\color{RDFKeyword}-}}}1
    {|}{{{\color{RDFKeyword}|}}}1
    {>}{{{\color{RDFKeyword}>}}}1,
  classoffset=0
}


\maketitle

\section{Introduction} \label{sec:intro}
The need for validating RDF data has led to the development of two shape-based languages: ShEx and SHACL. 
 While both languages share the goal of validating RDF data, they differ in their approach and implementation.

 ShEx~\cite{PGS14} was created as a concise and human-readable language, inspired by regular expressions and the idea of defining 
  a ShEx schema as a set of shapes that describe a kind of grammar for RDF data. 
  ShEx fully embraced the concept of recursive shapes, allowing for the definition of shapes that can refer to themselves~\cite{SBG15}. 
  In order to accommodate the combination of recursive shapes with negation, 
  the semantics of ShEx was defined in terms of stratified negation which is a well-defined and widely accepted approach 
  for handling recursion and negation in logic programming~\cite{SBG15}. 

  SHACL, on the other hand, was designed to provide a comprehensive framework for validating RDF data as a set of constraints that can be applied to RDF graphs. 
  While SHACL also allows for the definition of recursive shapes, the semantics of recursive shapes is left to the implementation of the different SHACL engines. 
  As a result, different SHACL engines may exhibit different behaviours in the presence of recursive shapes, 
  leading to inconsistencies and potential issues in the validation process.
 There have been several approaches to describe the SHACL semantics with recursive shapes, 
 including the use of fixed-point semantics and stratified negation
  but none of them has been adopted as part of the W3C recommendation. 
  Although SHACL 1.2 specification is currently being developed, it does not yet provide a definitive solution to this issue\footnote{\url{https://www.w3.org/TR/shacl12-core/\#shapes-recursion}}.

A formal analysis of different notions of recursion for shape-based languages was done in~\cite{recursive_shapes}.
 In that paper, the sheval tool was developed as a framework for testing and comparing the behaviour
  of different shape engines, specifically addressing recursion. 
 The purpose of sheval is to help practitioners and researchers identify inconsistencies between, 
  and potential issues with, 
  existing ShEx and SHACL validators when confronted with different test cases. 
  In particular, we focus on the behaviour of these validators when dealing with recursive shapes 
  and propose a test suite which covers different scenarios of recursion, 
  including the combination of recursion with negation.
 
 The main contribution of this paper is to present a description of the architecture of the sheval tool 
 and an extended test suite for recursive shapes which includes control tests for non-recursive shapes and tests for the duality proposition between \LFP{} and \GFP{} as well as a more in-depth analysis of the results. 

The \textsc{sheval} tool is available as an open-source project on 
 GitHub\footnote{\url{https://github.com/cogsl/sheval}}, 
 including the source code, 
 three test suites and
 the binaries of the engines employed in the comparison. 
 The tool can be run in Docker to help reproducibility of the results.
\section{Simple Shape Language and Recursive shapes} \label{sec:recursive_shapes}

\begin{definition}
Given a set of IRIs $I$, a set of blank-nodes $B$, and a set of literals $L$, an RDF graph $G$ is a set of triples $(s, p, o)$ where 
 $s \in I \cup B$ is the subject, $p \in I$ is the predicate, and $o \in I \cup B \cup L$ is the object. 
\end{definition}

\begin{example}
Figure~\ref{example_rdf} shows an RDF graph in Turtle notation. It represents a simple social network with nodes \lstinline|:a|, \lstinline|:b|, and \lstinline|:c| 
 whose names are \lstinline|"Alice"|, \lstinline|"Bob"| and \lstinline|"Carol"|, respectively,
 where \lstinline|:a| knows \lstinline|:c|, \lstinline|:b| knows \lstinline|:c| and \lstinline|:c| knows \lstinline|:a|.
Figure~\ref{example_rdf} shows a graphical representation of this same RDF graph.

\begin{figure}[t]
\centering
\begin{minipage}[t]{0.46\textwidth}
\vspace{0pt}
\begin{lstlisting}[language=Turtle]
prefix :       <http://example.org/>

:a :name "Alice" ;
   :knows :c .
:b :name "Bob" ;
   :knows :c .
:c :name "Carol" ;
   :knows :a     .
\end{lstlisting}
\end{minipage}%
\hfill
\begin{minipage}[t]{0.5\textwidth}
\vspace{0pt}
\begin{tikzpicture}[
  resource/.style={draw, ellipse, minimum width=12mm, minimum height=8mm},
  literal/.style={draw, rectangle, minimum width=16mm, minimum height=8mm},
  arr/.style={-{Latex[length=2mm]}, thick}
]

\node[resource] (a) {\texttt{:a}};
\node[resource, below=15mm of a] (b) {\texttt{:b}};
\node[resource] (c) at ($(a)!0.5!(b)+(30mm,0)$) {\texttt{:c}};

\node[literal, right=15mm of a, yshift=5mm] (alice) {Alice};
\node[literal, right=15mm of b, yshift=-5mm] (bob) {Bob};
\node[literal, right=15mm of c] (carol) {Carol};

\draw[arr] (a) -- node[pos=0.5, above, sloped] {\small \texttt{:name}} (alice);
\draw[arr] (b) -- node[pos=0.5, below, sloped] {\small \texttt{:name}} (bob);
\draw[arr] (c) -- node[above] {\small \texttt{:name}} (carol);

\draw[arr] (a) edge[bend left=20] node[pos=0.7, above, sloped] {\small \texttt{:knows}} (c);
\draw[arr] (c) edge[bend left=20] node[pos=0.5, above, sloped] {\small \texttt{:knows}} (a);

\draw[arr] (b) -- node[pos=0.4, above, sloped] {\small \texttt{:knows}} (c);

\end{tikzpicture}
\end{minipage}
\caption{Example RDF graph representing a simple social network}
\label{example_rdf}
\end{figure}
\end{example}

Following~\cite{recursive_shapes}, we define a simple shape language that can be used to describe the concept of recursive shapes. 

\begin{definition}
 Given a set of shape names $N$, 
 we define the set of shapes $S$ as the set of all possible \emph{shapes} $\varphi$ defined by the following syntax, 
 where $s \in N$ is a shape name, 
       $p \in I$ is a predicate, and 
       $\hasvalue$ is a boolean condition over nodes.

\begin{align*}
  \varphi
  \gDef
  & \bot \! 
  \gMid \top \!
  \gMid \hasvalue \!
  \gMid s \!
  \gMid \neg \varphi \!
  \gMid  \varphi \lor \varphi \!
  \gMid \varphi \land \varphi \!
  \gMid \exists p. \varphi \!
  \gMid \forall p. \varphi \!
  \gEnd
\end{align*}

An informal semantics of the language features is defined as follows\footnote{For the complete formal semantics, we refer to~\cite{recursive_shapes}}:
\begin{itemize}
  \item $\bot$ represents an empty shape, which is not satisfied by any node and $\top$ the universal shape, which is satisfied by all nodes.
  \item $\hasvalue$ is satisfied if the node satisfies $\hasvalue$. 
   In this paper will use only two conditions: 
    $\in String$ which is satisfied if the node is a String literal, and 
    $\test(n)$ which is satisfied if the node is $n$.
  \item $s$ is a reference to another shape. A node $n$ conforms to $s$ when it satisfies the definition of $s$. Given that the language can have recursive definitions, the precise semantics in case of recursion will need shape assignments which will be introduced later. 
  \item $\neg \varphi$ is the negation of a shape, satisfied by a node if it does not satisfy $\varphi$.
  \item $\varphi_1 \lor \varphi_2$ is satisfied if a node satisfies at least one of $\varphi_1$ or $\varphi_2$.
  \item $\varphi_1 \land \varphi_2$ is satisfied if a node satisfies both shapes.
  \item $\exists p. \varphi$ is satisfied by a node if there is an outgoing predicate $p$ to a node that satisfies $\varphi$.
  \item $\forall p. \varphi$ is satisfied if all outgoing edges with predicate $p$ lead to nodes that satisfy $\varphi$.
\end{itemize}

A \emph{simple shape language catalog} 
$\cat: N\rightarrow{}S$ is a partial function mapping shape names to shapes. 
 We assume that all shapes used in the image of the function are defined.

\end{definition}

\begin{example} \label{example_catalog}
We can define a simple shape language catalog $\cat$ that describes the social network in Figure~\ref{example_rdf} as follows:

$\begin{array}[t]{cl} 
\cat = \{ & user \mapsto person\ \land \exists\,name.(\in{} String)\ \land \exists\ knows.\top, \\
    & person \mapsto user\ \land \forall\,knows.person\ \} \\
\end{array}$


The catalog defines two mutually recursive shapes: $user$ and $person$. A node conforms to shape $user$ if it conforms to $person$ and has at least one property $name$ whose value is a literal string and at least another property $knows$. A node conforms to the shape $person$ if it also conforms to $user$ and all values of property $knows$ conform to $person$.
\end{example}

\begin{definition}
A shape assignment $\alpha$ for a shapes catalog $\cat$ and graph $\graph$ 
 is a relation $\alpha \subseteq (I \cup B \cup L) \times N$ 
 that associates nodes in the graph with shape names in the schema.
 A shape assignment $\alpha$ is said to be \emph{valid} if and only if for every 
  $(n, s) \in \alpha$, the node $n$ satisfies the shape $\cat(s)$ 
  according to the semantics of the simple shape language.
\end{definition}

Shape assignments are necessary to indicate which nodes in the graph we want to validate against which shapes, 
 in ShEx they are called shape maps, while in SHACL they can be defined with target declarations. 
 Shape assignments can also be used to identify the result of the validation 
and during the validation with recursive shapes, 
they are used to define the semantics of the validation.

When dealing with recursive shapes, it is possible that there is more than one valid shape assignment for a given schema and graph. 

\begin{example}
Given the catalog $\cat$ from example~\ref{example_catalog} 
 and the graph $\graph$ from example~\ref{example_rdf}, 
 notice that there is one possibility to assign the shape 
 \lstinline|user| to all nodes in the graph,
 another possibility would be to assign the shape \lstinline|user| to nodes \lstinline|:a| and \lstinline|:c|,
 and a third possibility would be to assign no shape to any node in the graph. 
 Hence, we have three valid shape assignments $\alpha_1, \alpha_2, \alpha_3$ :

$\begin{array}[t]{l} 
\alpha_1 = \{ (\text{:a}, user), (\text{:b}, user), (\text{:c}, user), (\text{:a}, person), (\text{:b}, person), (\text{:c}, person) \} \\
\alpha_2 = \{ (\text{:a}, user), (\text{:c}, user), (\text{:a}, person), (\text{:c}, person)  \} \\
\alpha_3 = \{ \} \\
\end{array}$
\end{example}

There are three common semantics for recursive shapes: least fixed point (LFP), 
 greatest fixed point (GFP), and supported model semantics (SMS). 
 LFP assumes that a node does not conform to a shape unless it satisfies the shape's constraints, 
 in the previous example, LFP is $\alpha_3$. 
 GFP assumes that a node conforms to a shape unless it violates the shape's constraints so in the previous example, GFP is $\alpha_1$. 
 The SMS looks for any self-supporting interpretation, not necessarily the least or the greatest one, and in the previous example, 
  the SMS is the set $\{\alpha_1, \alpha_2, \alpha_3\}$.
 In the case of SMS, a validator is considered to use brave Supported Model Semantics (bSMS) 
  if it checks that its result is any of the possible shape assignments, 
 and it is considered cautious Supported Model Semantics (cSMS) 
 if it checks all possible shape assignments.

For more details about LFP and GFP semantics, 
 \cite{recursive_shapes} presents some results that show that there is a significant fragment of ShEx with GFP and of SHACL with LFP 
  that have the same expressive power, with possible translations between them.

\section{ShEx, SHACL and recursive shapes} \label{sec:diff_shex_shacl_recursion}


Although both ShEx and SHACL are shape-based languages for validating RDF data, 
 they differ in their approach and implementation. 
In this section, we will discuss some of the key differences between ShEx and SHACL engines 
 and how they handle recursive shapes.

Figure~\ref{example_shapes} 
 presents both the ShEx schema and the equivalent SHACL shapes graph 
 corresponding to Example~\ref{example_catalog}. 
 In the case of ShEx, we added a second declaration for \lstinline|:name| 
 to capture the semantics of the Simple Shape Language in which the property 
 declarations are open.
 Similarly, in the case of SHACL, we use \lstinline|sh:qualifiedValueShape| 
 to capture the semantics of the Simple Shape Language,
 which would allow extra values for \lstinline|:name|.

\begin{figure}[t]
\centering
\begin{minipage}[t]{0.48\linewidth}
\begin{lstlisting}[language=ShExC, aboveskip=0pt]
:User @:Person AND {
  :name  xsd:string    ;
  :name  .           * ;
  :knows .           + ;
}

:Person @:User AND {
  :knows @:Person    * 
}
\end{lstlisting}
\end{minipage}%
\hfill
\begin{minipage}[t]{0.48\linewidth}
\begin{lstlisting}[language=SHACL, aboveskip=0pt]
:User a sh:NodeShape ;
  sh:and ( 
      :Person
      [ sh:property 
          [ sh:path :name ;
            sh:qualifiedValueShape [ 
              sh:datatype xsd:string 
            ];
            sh:qualifiedMinCount 1
          ] ] 
      [ sh:property 
          [ sh:path :knows ;
            sh:minCount 1
          ] ] ) .

:Person a sh:NodeShape ;
    sh:and ( 
      :User 
      [ sh:property 
          [ sh:path :knows ;
            sh:node :Person ;
          ] ;
      ]) .
\end{lstlisting}
\end{minipage}
\caption{ShEx schema, SHACL shapes graph corresponding to example~\ref{example_catalog}}
\label{example_shapes}
\end{figure}

 ShEx validators usually have three inputs, an RDF graph, a ShEx schema 
 and a shape map that describes which nodes should be checked against which shapes,
 and they return a result map with the values that conform (or not) to the expected shapes. 
 The shape maps are similar to shape assignments that indicate which nodes conform 
 with which shapes.

 \begin{example}
 A shape map to validate the nodes \lstinline|:a|, \lstinline|:b| and \lstinline|:c| against the shape \lstinline|:User| is \lstinline|:a@:User, :b@:User, :c@:User|.
 Applying that shape map to the RDF graph in example~\ref{example_rdf}, 
 the basic result shape map obtained in ShEx is 
 \lstinline|:a@:User, :b@:User, :c@:User|, 
 indicating that all three nodes conform to the shape \lstinline|:User| 
 (in practice, ShEx validators return the result using different syntaxes 
  like JSON or a table with more information about the evidences for conformance/nonconformance).
\end{example}

In the previous example, the conformance is based on the fact that 
 ShEx specification~\cite{PBGK19} which is based on~\cite{BGP17} 
 explicitly declares GFP semantics to evaluate recursive shapes.

SHACL engines have as input an RDF graph and a SHACL shapes graph, 
 and they return a validation report that indicates 
 if the RDF graph conforms to the SHACL shapes graph or if there are violations. 
 SHACL shapes can also include target declarations which act like shape assignments.
 After running a SHACL validator with an RDF graph, a shapes graph and target declarations, it is expected to return a validation report which returns \lstinline|conforms=true| if all the nodes selected by the target declarations conform to the expected shapes or \lstinline|conforms=false| and a list violation errors if some nodes don't conform to their expected shapes.
 The SHACL recommendation  does not specify a specific semantics for recursive shapes, 
 so the behaviour of SHACL engines confronted with recursive shapes is not uniform. 

\begin{example}
If we run the previous example with the SHACL engine \texttt{TopQuadrant SHACL API}~\cite{tq}, 
  the result obtained is the following validation report which indicates that it considers that the nodes \lstinline|:a| and \lstinline|:b| don't conform to the shape \lstinline|:Person|:
{\small
\begin{lstlisting}[language=SHACL]
[ rdf:type     sh:ValidationReport;
  sh:conforms  false;
  sh:result [ a sh:ValidationResult;
    sh:focusNode :a;
    sh:resultMessage 
      "Value must have all of the following shapes: :User, _:1";
    sh:resultSeverity sh:Violation;
    sh:sourceConstraintComponent sh:AndConstraintComponent;
    sh:sourceShape :Person;
    sh:value :a
  ];
  sh:result [ a sh:ValidationResult;
    sh:focusNode :b;
    sh:resultMessage 
      "Value must have all of the following shapes: :User, _:2";
    sh:resultSeverity sh:Violation;
    sh:sourceConstraintComponent  sh:AndConstraintComponent;
    sh:sourceShape :Person;
    sh:value :b
   ]
  . . . # Similar for node :c 
] .
\end{lstlisting}
}

By contrast, if we run the example with the Apache Jena SHACL engine~\cite{jenacl}  
 the result obtained is the following validation report:
{\small\begin{lstlisting}[language=SHACL]
[ rdf:type     sh:ValidationReport;
  sh:conforms  true;
] .
\end{lstlisting}
}
There are other possible behaviours: 
rudof~\cite{rudof} 
 generates an error indicating that it doesn't support recursive shapes, 
pySHACL~\cite{pySHACL} 
  generates a warning but continues the processing trying to validate and crashes with the message \lstinline|"Validation path too deep!"|  
  and SHACL-S~\cite{shacls} enters in an infinite loop without any warning.
\end{example}

Given that the SHACL 1.0 recommendation left handling of recursive shapes undefined, 
all the behaviours presented in the previous example are not breaking the recommendation. 
However, this situation leads to inconsistencies and lack of interoperability 
between different SHACL implementations and users' expectations. 
In the next section we present a tool and a set of tests that can be used to 
compare the behaviour of different shapes technologies 
when confronted with recursive definitions.

\section{sheval architecture}

\textsc{sheval} is a command-line framework for running the same shape-validation
test suite against different \emph{engines} that implement ShEx or SHACL. 
It also compares the obtained results against the 
 expected semantics that a suite may prescribe. 
Its design follows a single guiding principle: everything that is
 \emph{specific to one validation technology} (how to invoke it, how to parse
its output, how it signals a crash) is isolated behind a small, uniform
\texttt{Runner} interface, 
while everything that is \emph{common to all of them} 
(test orchestration, result classification, semantics comparison,
report generation) is implemented exactly once. 
Figure~\ref{fig:architecture}
gives an overview of the resulting module structure.

\paragraph{Test suites as data.} 
A test suite is a directory described by a \texttt{manifest.yaml} file, 
listing the RDF/SHACL/ShEx source folders, 
the implementations to run per engine,
and the test cases themselves. 
Each test case declares 
a data graph, a schema, and
the \emph{nodes} and \emph{shapes} or \emph{node--shape pairs} to validate,
plus an optional \texttt{expected\_results} block declaring the verdict
prescribed by one or more semantics (the greatest/least fixed points,
GFP/LFP, and the brave/cautious supported model semantics, bSMS/cSMS) as sets of
pairs expected to pass or fail. 
Because this is pure data, adding a test case
or a whole new suite never touches the code.

\paragraph{CLI and orchestration.} The entry point 
exposes four subcommands: 
\texttt{test} runs a full suite; 
\texttt{shacl}/
\texttt{shex} invoke a single technology directly on a data/schema pair; 
and \texttt{check} runs a technology's raw binary. 
For \texttt{test}, the \emph{orchestrator} loads the manifest and, 
 per test case, enriches the declared nodes/shapes/pairs 
 with full IRIs and delegates to \texttt{graph\_utils} 
 to materialize each engine's expected input: 
 a merged RDF graph with \texttt{sh:targetNode} declarations for SHACL, 
 or a ShapeMap file for ShEx. 
 It then dispatches these inputs to every requested technology
 and collects the results.

\paragraph{The Runner framework.} Each technology is wrapped by a
\texttt{Runner} subclass (e.g.\ \texttt{PyshaclRunner}, \texttt{JenaShaclRunner}, etc), 
 grouped under a \texttt{SHACLRunner} or \texttt{ShExRunner} marker class. 
 A concrete runner supplies a command-line \emph{template} for 
 its binary and implements two hooks: \texttt{execute}, 
  which fills the template, invokes the binary through the shared \texttt{commands} module, 
  and hands the output to \texttt{analysis} for parsing into a common
  \texttt{\{conforms, successes, failures, message\}} shape; 
  and an optional \texttt{classify\_error}, 
  recognizing that engine's own failure signatures
  (e.g.\ rudof's ``Dependency graph has cycles'', pySHACL's ``Validation path
too deep!''), etc. 
 The base class's \texttt{run} wraps every call in a common
 exception handler, so one technology crashing never aborts the suite.

\paragraph{Command execution and error classification.} 
Command execution is provided by a \texttt{commands} module,
which is in charge of spawning subprocesses: 
 it runs a binary under a timeout and returns a 
  \texttt{RunOutcome} with exit status, stderr and return code.
 \texttt{error\_classification} then maps a failed run to a recognized
  failure mode --- \texttt{Timeout}, \texttt{Crashed},
\texttt{CyclesDetected(Crashed/Conformant/NonConformant)},
\texttt{NonStratified(Crashed)} --- first via the runner's own
\texttt{classify\_error} hook, then via generic, engine-agnostic crash
signatures. 
This keeps the cross-cutting control flow in one place while
each engine contributes only the patterns it actually produces.

\paragraph{Semantics matching and reporting.} 
The \texttt{semantics} module compares an
engine's reported successes/failures against every semantics declared in a
test's \texttt{expected\_results}, 
recording agreement with at least one
(brave) or all (cautious) of the candidate models. 
The annotated results are
finally handed to one or more exporters: 
plain YAML/CSV, or a LaTeX table like the one employed for Table~\ref{tab:sheval-results}.

\begin{figure}[t]
  \centering
  \resizebox{0.85\textwidth}{!}{%
  \begin{tikzpicture}[
    module/.style={rectangle, draw, thick, rounded corners=2pt, fill=blue!8,
      minimum width=4.4cm, minimum height=1cm, align=center, font=\footnotesize, inner sep=2pt},
    data/.style={rectangle, draw, thick, fill=yellow!20,
      minimum width=4.2cm, minimum height=1cm, align=center, font=\footnotesize, inner sep=2pt},
    ext/.style={rectangle, draw, thick, dashed, fill=gray!12,
      minimum width=4.6cm, minimum height=1.2cm, align=center, font=\footnotesize, inner sep=2pt},
    tech/.style={rectangle, draw, fill=blue!4,
      minimum width=2.1cm, minimum height=0.7cm, align=center, font=\scriptsize, inner sep=1pt},
    grp/.style={rectangle, draw, dashed, rounded corners=2pt, inner sep=7pt},
    arr/.style={-{Latex[length=2.2mm]}, thick},
    tarr/.style={-{Latex[length=1.6mm]}},
    lbl/.style={font=\tiny, fill=white, inner sep=1pt},
    dots/.style={rectangle, inner sep=7pt},
    ]

    \node[data] (manifest) at (6.5,13.6)  {Test Suite\\Manifest\\\texttt{manifest.yaml}};

    \node[module] (orch) at (6.5,11.9) {Orchestrator};
    \node[module] (prep) at (6.5,10.5) {Input Preparation};

    \node[module] (runnerbase) at (6.5,8.7) {Runner};
    \node[module, minimum width=3.6cm] (shaclrunner) at (2.0,7.1)  {SHACLRunner};
    \node[module, minimum width=3.6cm] (shexrunner) at (11.0,7.1) {ShExRunner};

    \node[tech] (pyshacl)      at (-0.1,5.5) {pySHACL};
    \node[dots] (dots_shacl)   at (1.9,5.5) {...};
    \node[tech] (rudofshacl)   at (3.9,5.5) {rudof$_{SHACL}$};

    \node[tech] (rudofshex)    at ( 8.5,5.5) {rudof$_{ShEx}$};
    \node[tech] (shexs)        at (10.8,5.5) {ShEx-S};
    \node[tech] (jenashex)     at (13.1,5.5) {Jena$_{ShEx}$};

    \node[grp, fit=(runnerbase)(shaclrunner)(shexrunner)(pyshacl)(dots_shacl)(rudofshacl)(rudofshex)(shexs)(jenashex),
      label={[font=\small\bfseries, xshift=-2.7cm]above:Runner Framework}] (grpRunner) {};

    \node[module] (cmdexec) at (6.5,3.4) {Command Execution};
    \node[ext, minimum height=1.6cm] (extengines) at (15.2,3.4) {External Validator\\Engines\\(native / JVM / Python\\processes and CLIs)};

    \node[module] (analysis) at (3.7,1.1) {Result Analysis};
    \node[module, minimum height=1.3cm] (errclass) at (9.3,1.1) {Error Classification};
    \node[grp, fit=(analysis)(errclass), label={[font=\small\bfseries, xshift=-2.7cm]above:Result Processing}] (grpResult) {};

    \node[module] (semantics) at (6.5,-1.0) {Semantics Matching};
    \node[module, minimum width=6.4cm] (export) at (6.5,-2.9)
      {Exporters: YAML / CSV / LaTeX};

    \draw[arr] (manifest.south) -- (orch);
    \draw[arr] (orch) -- (prep);
    \draw[arr] (prep) -- (grpRunner.north);

    \draw[-{Triangle[open,length=8pt,width=8pt]}] (shexrunner) -- (runnerbase);
    \draw[-{Triangle[open,length=8pt,width=8pt]}] (shaclrunner) -- (runnerbase);

    \foreach \t in {pyshacl, rudofshacl}
      \draw[-{Triangle[open,length=8pt,width=8pt]}] (\t) -- (shaclrunner) ;
    \foreach \t in {rudofshex, shexs, jenashex}
      \draw[-{Triangle[open,length=8pt,width=8pt]}] (\t) -- (shexrunner) ;

    \draw[arr] (grpRunner.south) -- (cmdexec.north);
    \draw[arr] ($(cmdexec.east)+(0,0.2)$) -- node[lbl, above] {spawn} ($(extengines.west)+(0,0.2)$);
    \draw[arr] ($(extengines.west)+(0,-0.2)$) -- node[lbl, below] {stdout / stderr / exit code} ($(cmdexec.east)+(0,-0.2)$);

    \draw[arr] (cmdexec.south) -- node[lbl, pos=0.25] {RunOutcome} (grpResult.north);

    \draw[arr] (grpResult.south) -- (semantics.north);
    \draw[arr] (manifest.west) -- ++(-6.5,0) |- node[lbl, pos=0.85] {expected\_results} (semantics.west);
    \draw[arr] (semantics) -- node[lbl] {annotated results} (export);

  \end{tikzpicture}%
  }
  \caption{Main modules of \textsc{sheval}: the YAML manifest drives the orchestrator, which prepares
    engine-specific inputs; runners invoke external validation
    engines  whose output is parsed,
    classified and checked against the expected results and exported.}
  \label{fig:architecture}
\end{figure}





  




  

The repository includes a Docker-based setup with pinned dependencies 
 and bundled validator binaries for pySHACL, SHACL-S, rudof (ShEx and SHACL), Apache Jena (ShEx and SHACL), SHACL-TQ and ShEx-S, 
 enabling reproducible cross-engine comparisons\footnote{At the time of writing, the repository contains pySHACL version 0.30.1, SHACL-S version 0.1.87, rudof version 0.3.18, Apache Jena 5.3.0, SHACL-TQ version 1.4.4 and ShEx-s version 0.2.34}. 
Adding a new validator only requires to analyze and capture the template of the command required to run the validator and how it handles its output and potential warnings or errors, 
configure the necessary dependencies in Docker and 
implement a subclass of either \lstinline|ShExRunner| or \lstinline|SHACLRunner|.

\section{Recursive shapes test suite} \label{sec:recursion}

We developed a test suite to compare the behaviour of different shape engines when confronted with recursive shape definitions. 
The test suite aims to define simple tests that check the different semantic possibilities of recursive shapes. 
The test cases fall into two categories: 
 separation and feature tests. 
Separation tests attempt to distinguish whether an engine uses 
 \LFP, 
 \GFP, 
 brave SMS, or 
 cautious SMS. 
Feature tests check some specific features of the engines. 
 Each test is defined as $t_i  = \langle \graph, \cat, \sel, \alpha_{\LFP}, \alpha_{\GFP}, \alpha_1, \ldots, \alpha_n \rangle$
 where $\graph$ is an RDF graph,
 $\cat$ is a shape catalog, 
 and $\sel$ is a selector,
 where $\alpha_{\LFP}$ is the expected shape assignment with least fixed-point semantics, 
 $\alpha_{\GFP}$ with greatest fixed-point semantics, and 
 $\alpha_1, \ldots, \alpha_n$ are the remaining expected shape assignments in case of multiple valid interpretations in supported model semantics. 
An assignment $\alpha$ is {\pass green}, if $\graph$ conforms to $(\cat, \sel)$ under $\alpha$,  
and  {\fail brown}, if it does not.
The shape catalogs are defined using the abstract syntax of the simple shape languages. 
The conversion of those catalogs to ShEx and SHACL is presented in the github repository.

\vspace{-1mm}
\paragraph{Basic separation tests.}
These four tests are designed to distinguish a validator that employs $\LFP$ semantics from one that employs $\GFP$, using minimal examples. 

{ 
\vspace{-1mm}

\begin{itemize}
\item \texttt{bsep1}: 
$\begin{array}[t]{l}
\graph = \{ (a,p,a) \}\,,\ \mathcal{C} =  \{ \decl{s}{\exists p . s} \}\,,\ 
\sel = \{ (a,s)\}\, \\
\fail{\alpha_{\LFP} = \emptyset}\,,\ \pass{\alpha_{\GFP} = \{(a,s)\}}\,.
\end{array}$

\item \testcode{bsep2}: 
$\begin{array}[t]{l} 
\graph = \{(a,p,a), (b,p,b) \}\text{, }
\mathcal{C} =  \{\decl{s}{\exists p . s} \}\text{, }
 \sel = \{ (a,s), (b,s)\}\,,\ \\
 {\fail \alpha_{\LFP}=\emptyset}\,,\ 
 {\pass \alpha_{\GFP}=\{(a, s),(b,s)\}}\,,\ \\
 {\fail \alpha_1=\{(a,s)\}}\,,\ 
 {\fail \alpha_2 =\{(b,s)\}}\,.   
\end{array}$

\item  \testcode{bsep3}: 
$
\begin{array}[t]{l}

%
\graph = \{(a,p,c), (b,p,c) \}\text{, }
\mathcal{C} =  \{\decl{s}{s' \land \exists p. \top},\ \decl{s'}{s \land \exists p. \top} \}, \sel = \{ (a,s), (b,s)\}\,,\\
{\fail \alpha_{\LFP} = \emptyset}\,,\ 
{\pass \alpha_{\GFP}  = \{(a,s), (b,s), (a,s'), (b,s')\}}\,,\ \\
{\fail \alpha_1 = \{(a,s), (a,s')\}}\,,\ 
{\fail \alpha_2 = \{(b,s), (b,s')\}}\,.  
\end{array}$
\item \testcode{bsep4}:  
$
\begin{array}[t]{l}
\graph \text{ as in \testcode{bsep2}, }\mathcal{C} = \{\decl{s}{\exists p.s}, \decl{s'}{\neg s} \},\ 
\sel = \{ (a,s), (b,s)\}\,,\\
{\fail \alpha_{\LFP} = \{(a, s'), (b,s')\}}\,,\ 
{\pass \alpha_{\GFP} =  \{(a, s), (b, s) \}}\,, \\    
{\fail \alpha_1 = \{(a,s), (b, s') \}}\,,\ 
{\fail \alpha_2 = \{(b,s), (a, s') \}}\,. 
\end{array}$      
\end{itemize}
}
\paragraph{Reachability separation tests.}
The next four tests check reachability and its dual property of safety by 
 checking them on two dual shape names $r$ and $s$, 
 in four different scenarios. 
 All 4 tests use the same graph: 
 $\graph=\{(a,p,d), (b,p,a), (b,p,c), (c,p,d), (d,p,c) \}$ 

{ 

\begin{itemize}
\item \testcode{reach1}:  
$
\begin{array}[t]{l} 
\mathcal{C} =  \{\decl{r}{\test(a) \lor \exists p.r}  \}, 
\sel = \{ (a,r), (b,r), (c,r), (d,r)\}\,,\\
{\fail \alpha_{\LFP}= \{(a,r), (b,r)\}} \,,\\
{\pass \alpha_{\GFP}=\{(a,r),(b,r),(c,r),(d,r)\}}\,. \\ 
\end{array}$

\item \testcode{reach2}: 
$\begin{array}[t]{l}
\mathcal{C} =  \{\decl{s}{\neg \test(a) \land \forall p. s},\ \decl{r}{\neg s} \}, 
\sel = \{ (c,r), (d,r) \}\,,\\
{\pass \alpha_{\LFP}=\{(a,r),(b,r),(c,r),(d,r)\}}\,, \\
{\fail \alpha_{\GFP}=\{(a,r), (b,r),(c,s),(d,s)\}}\,. 
\end{array}$

\item \testcode{safe1}:  
$\begin{array}[t]{l}
\mathcal{C} =  \{\decl{s}{\neg \test(a) \land \forall p. s} \}\text{, }
\sel = \{ (c,s), (d,s) \}\,,\\
{\fail \alpha_{\LFP}=\emptyset}\,,\ 
{\pass \alpha_{\GFP}=\{(c,s)(d,s) \}}\,.     
\end{array}$

\item \testcode{safe2}:  
$
\begin{array}[t]{l}
  \mathcal{C} =  \{\decl{r}{\test(a) \lor \exists p. r},\ \decl{s}{\neg r} \},
  \sel = \{ (c,r), (d,r)\}\,,\\
{\fail \alpha_{\LFP} = \{(a,r), (b,r),(c,s),(d,s)\}}\,,\ \\
{\pass \alpha_{\GFP} = \{(a,r),(b,r),(c,r),(d,r)\}}\,.
\end{array}$
\end{itemize}

}
\paragraph{Feature tests} These target more specific properties of the validator, as we will explain after introducing them.
{ 
\begin{itemize}
\item \texttt{nstrat1} (non-stratified negation without cyclic data): \\
$ \graph = \{ (a,p,b), (b,p,c) ,(c,p,d), (d,p,e)  \}, 
  \mathcal{C}=\{ \decl{s}{\exists p.\neg s}  \}$,  
  $\sel = \{ (a,s), (b,s), (c,s), (d,s), (e,s) \}
$\\
\emph{Expected result}: A validation engine can reject the catalog for non-stratified shapes. 

\item \testcode{nstrat2} (non-stratified negation with cyclic data): \\
$\graph = \{ (a,p,b), (b,p,a)  \}, 
\mathcal{C} =\{ \decl{s}{\exists p.\neg s} \}$, 
$\sel = \{ (a,s), (b,s)\}$\\
\emph{Expected result}:  A validation engine can reject the catalog for non-stratified shapes.

\item \texttt{cons1} (consistency): 
$\graph = \{ (a,p,a)\}$, 
$\cat=\{ \decl{s}{\exists p.s'}, \decl{s'}{\neg s} \}$, 
$\sel = \{ (a,s) , (a,s') \}$\\
\emph{Expected result}: The engine should reject because $a$ can't conform to $s$ and $s'$ at the same time.

\item \testcode{cons2} (consistency): $\graph \text{ as in \testcode{nstrat2}},\ \cat \text{ as in \testcode{cons1}}$,
$\sel = \{ (a,s), (a,s'),(b,s),(b,s') \}$\\
\emph{Expected result}: The engine should reject because $a$ and $b$ can't conform to $s$ and $s'$ at the same time.

\item \texttt{fresh} (fresh constant support):\\ $\graph  = \{  (a,p,b), \allowbreak (b,p,c), \allowbreak (c,p,a) \}$, 
$\mathcal{C} = \{ \decl{s}{\top} \}$, \\ 
$\sel = \{ (d,s) \}$\\
\emph{Expected result}: The engine can accept because any value conforms to shape $s$, although as $d$ is not part of the graph, it could also reject it.

\end{itemize}
}

The \testcode{nstrat1} and \testcode{nstrat2} tests contain a combination of negation and recursion which is non-stratified. 
 The tests are included in the test suite  
 to check how engines behave on such inputs although the proposed semantics doesn't support non-stratified catalogs. 
\testcode{cons1} and \testcode{cons2} check for 
 \emph{logical consistency}. 
 We have a catalog with two shapes, $s$ and $s'$, 
  where $s'$ is defined as the negation of $s$. 
  Then we ask in the shape map that both $s$ and $s'$ 
  be assigned to the same node. 
  Since no such assignment is possible while being consistent with 
  the catalog's semantics, a consistent validator should reject this possibility.
Finally, the test \testcode{fresh} has a shape catalog with a shape assignment 
 that is trivially satisfied, using the shape $\top$. 
In the shape map, we require that this shape is satisfied by a ``fresh node'', 
 that is, a node that is not featured in the graph. 
This reflects something that both SHACL and ShEx permit: 
 selecting nodes outside the input graph. 
  
\paragraph{Duality tests} These tests check the duality proposition presented 
 in ~\cite{recursive_shapes} between \LFP{} and \GFP{}. 
\texttt{dual1} uses a catalog which contains a negation, a conjunction and 
a universal quantifier, 
while \texttt{dual2} contains the dual shape that swaps negation by a direct test, 
 conjunction by disjunction and $\forall$ by $\exists$. 
 The resulting shape assignments for \LFP{} and \GFP{} are complementary. 

{ 
\begin{itemize}
\item \texttt{dual1} (Combines existential and basic test): \\
$\begin{array}[t]{l}
\graph = \{ (a,p,b), (b,p,a) ,(b,q,c), (d,p,d) \}\,,
\mathcal{C}=\{\decl{s}{\lnot \test(b) \land \forall p. s}\,\}\,, \sel = \{ (d,s) \}\,,\\
{\fail \alpha_{\LFP}=\{(c,s)\}}\,, \\
{\pass \alpha_{\GFP}=\{(c,s), (d,s)\}}\,. 
\end{array}$

\item \texttt{dual2} (dual shape of \texttt{dual1} which contains a universal quantifier and basic test): \\
$\begin{array}[t]{l}
\graph = \text{as in \texttt{dual1}} \,,
\mathcal{C}=\{\decl{s}{\test(b) \lor \exists p. s}\}\,, \sel = \{ (d,s) \}\,,\\
{\fail \alpha_{\LFP}=\{(a,s),(b,s)\}}\,, \\
{\pass \alpha_{\GFP}=\{(a,s),(b,s),(d,s)\}}\,. 
\end{array}$

\end{itemize}
}

\paragraph{Non-recursive tests} We added two control tests \texttt{norec1} and \texttt{norec2} to check that non-recursive shapes behave as expected.  
In both cases, the shape assignments for \LFP{}, \GFP{}, bSMS and cSMS are equivalent. 

{ 
\begin{itemize}
\item \texttt{norec1} (basic non-recursive test): \\
$\begin{array}[t]{l}
\graph = \{ (a,p,a) \}\,,
\mathcal{C}=\{\decl{s}{\exists p. s'}\,\decl{s'}{test(a)}\}\,,
\sel = \{ (a,s) \}\,,\\
{\pass \alpha_{\LFP}=\alpha_{\GFP}=\{(a,s), (a,s')\}}\, \\
\end{array}$

\item \texttt{norec2} (non-recursive test expecting non-conformance): \\
$\begin{array}[t]{l}
\graph =  \{ (a,p,a), (b,p,b) \}\,,
\mathcal{C}=\{ \decl{s}{\exists p. \lnot s'}\,\decl{s'}{test(a)} \} \,,
\sel = \{ (a,s) \}\,,\\
{\fail \alpha_{\LFP}=\alpha_{\GFP}=\{(b,s), (a,s')\}}\, \\
\end{array}$

\end{itemize}
}

\providecommand{\shevalPass}{{\pass\scalebox{2}{$\text{\textbullet}$}}}
\providecommand{\shevalFail}{{\fail{$\blacksquare$}}}
\providecommand{\shevalNA}{--}
\providecommand{\shevalError}{{\error\textsf{X}}}
\providecommand{\shevalCyclesDetectedStopped}{{\stopped\textsf{C}}}
\providecommand{\shevalNonStratifiedStopped}{{\stopped\textsf{N}}}
\providecommand{\shevalTimeout}{{\error\textsf{T}}}
\providecommand{\shevalCrashed}{{\error\textsf{X}}}
\providecommand{\shevalCyclesDetectedCrashed}{{\error\textsf{C!}}}
\providecommand{\shevalNonStratifiedCrashed}{{\error\textsf{N!}}}
\providecommand{\shevalCyclesDetectedConformant}{{\pass\textsf{C+}}}
\providecommand{\shevalCyclesDetectedNonConformant}{{\fail\textsf{C-}}}

\providecommand{\shevalTechRudof}{rudof}
\providecommand{\shevalTechShexS}{ShEx-S}
\providecommand{\shevalTechJenaShex}{Jena}
\providecommand{\shevalTechPyshacl}{pySHACL}
\providecommand{\shevalTechShaclS}{SHACL-S}
\providecommand{\shevalTechJenaShacl}{Jena}
\providecommand{\shevalTechShaclTq}{SHACL TQ}
\providecommand{\shevalTechRudofNone}{rudof no cycles}
\providecommand{\shevalTechRudofCautious}{rudof LFP}
\providecommand{\shevalTechRudofBrave}{rudof GFP}

Table~\ref{tab:sheval-results} represents the results of running sheval using example~\ref{example_shapes} and the recursive shapes test suite. 

\begin{table}
\centering
\setlength{\tabcolsep}{5pt} 
\renewcommand{\arraystretch}{1.0}
\begin{tabular}{l|ccc|ccccccc|cccc}
Test & 
 \multicolumn{1}{c}{\adjustbox{angle=45,lap=\width}{\shevalTechRudof}} & 
 \multicolumn{1}{c}{\adjustbox{angle=45,lap=\width}{\shevalTechShexS}} & 
 \multicolumn{1}{c}{\adjustbox{angle=45,lap=\width}{\shevalTechJenaShex}} & 
 \multicolumn{1}{c}{\adjustbox{angle=45,lap=\width}{\shevalTechPyshacl}} & 
 \multicolumn{1}{c}{\adjustbox{angle=45,lap=\width}{\shevalTechShaclS}} & 
 \multicolumn{1}{c}{\adjustbox{angle=45,lap=\width}{\shevalTechJenaShacl}} & 
 \multicolumn{1}{c}{\adjustbox{angle=45,lap=\width}{\shevalTechShaclTq}} & 
 \multicolumn{1}{c}{\adjustbox{angle=45,lap=\width}{\shevalTechRudofNone}} & 
 \multicolumn{1}{c}{\adjustbox{angle=45,lap=\width}{\shevalTechRudofCautious}} & 
 \multicolumn{1}{c}{\adjustbox{angle=45,lap=\width}{\shevalTechRudofBrave}} & 
 \multicolumn{1}{c}{\adjustbox{angle=45,lap=\width}{GFP}} & 
 \multicolumn{1}{c}{\adjustbox{angle=45,lap=\width}{LFP}} & 
 \multicolumn{1}{c}{\adjustbox{angle=45,lap=\width}{bSMS}} & 
 \multicolumn{1}{c}{\adjustbox{angle=45,lap=\width}{cSMS}} \\
\hline
example~\ref{example_catalog} & \shevalPass & \shevalPass & \shevalPass & \shevalCyclesDetectedCrashed & \shevalTimeout & \shevalCyclesDetectedConformant & \shevalFail & \shevalCyclesDetectedStopped & \shevalFail & \shevalPass & \shevalPass & \shevalFail & \shevalPass & \shevalFail \\
\hline
bsep1 & \shevalPass & \shevalPass & \shevalPass & \shevalCyclesDetectedConformant & \shevalPass & \shevalCyclesDetectedConformant & \shevalPass & \shevalCyclesDetectedStopped & \shevalFail & \shevalPass & \shevalPass & \shevalFail & \shevalPass & \shevalFail \\
bsep2 & \shevalPass & \shevalPass & \shevalPass & \shevalCyclesDetectedConformant & \shevalPass & \shevalCyclesDetectedConformant & \shevalPass & \shevalCyclesDetectedStopped & \shevalFail & \shevalPass & \shevalPass & \shevalFail & \shevalPass & \shevalFail \\
bsep3 & \shevalPass & \shevalPass & \shevalPass & \shevalCyclesDetectedCrashed & \shevalTimeout & \shevalCyclesDetectedConformant & \shevalFail & \shevalCyclesDetectedStopped & \shevalFail & \shevalPass & \shevalPass & \shevalFail & \shevalPass & \shevalFail \\
bsep4 & \shevalPass & \shevalPass & \shevalPass & \shevalCyclesDetectedConformant & \shevalPass & \shevalCyclesDetectedConformant & \shevalPass & \shevalCyclesDetectedStopped & \shevalFail & \shevalPass & \shevalPass & \shevalFail & \shevalPass & \shevalFail \\
reach1 & \shevalPass & \shevalPass & \shevalPass & \shevalCyclesDetectedConformant & \shevalPass & \shevalCyclesDetectedConformant & \shevalFail & \shevalCyclesDetectedStopped & \shevalFail & \shevalPass & \shevalPass & \shevalFail & \shevalPass & \shevalFail \\
reach2 & \shevalFail & \shevalFail & \shevalFail & \shevalFail & \shevalFail & \shevalCyclesDetectedNonConformant & \shevalFail & \shevalCyclesDetectedStopped & \shevalPass & \shevalFail & \shevalFail & \shevalPass & \shevalPass & \shevalFail \\
safe1 & \shevalPass & \shevalPass & \shevalPass & \shevalCyclesDetectedConformant & \shevalPass & \shevalCyclesDetectedConformant & \shevalPass & \shevalCyclesDetectedStopped & \shevalFail & \shevalPass & \shevalPass & \shevalFail & \shevalPass & \shevalFail \\
safe2 & \shevalPass & \shevalPass & \shevalPass & \shevalCyclesDetectedConformant & \shevalPass & \shevalCyclesDetectedConformant & \shevalFail & \shevalCyclesDetectedStopped & \shevalFail & \shevalPass & \shevalPass & \shevalFail & \shevalPass & \shevalFail \\
\hline
nstrat1 & \shevalNonStratifiedStopped & \shevalNonStratifiedStopped & \shevalFail & \shevalFail & \shevalFail & \shevalCyclesDetectedNonConformant & \shevalFail & \shevalNonStratifiedStopped & \shevalNonStratifiedStopped & \shevalNonStratifiedStopped & \shevalNA & \shevalNA & \shevalNA & \shevalNA \\
nstrat2 & \shevalNonStratifiedStopped & \shevalNonStratifiedStopped & \shevalPass & \shevalFail & \shevalFail & \shevalCyclesDetectedNonConformant & \shevalFail & \shevalNonStratifiedStopped & \shevalNonStratifiedStopped & \shevalNonStratifiedStopped & \shevalNA & \shevalNA & \shevalNA & \shevalNA \\
cons1 & \shevalNonStratifiedStopped & \shevalNonStratifiedStopped & \shevalFail & \shevalFail & \shevalFail & \shevalCyclesDetectedNonConformant & \shevalFail & \shevalNonStratifiedStopped & \shevalNonStratifiedStopped & \shevalNonStratifiedStopped & \shevalNA & \shevalNA & \shevalNA & \shevalNA \\
cons2 & \shevalNonStratifiedStopped & \shevalNonStratifiedStopped & \shevalPass & \shevalFail & \shevalFail & \shevalCyclesDetectedNonConformant & \shevalFail & \shevalNonStratifiedStopped & \shevalNonStratifiedStopped & \shevalNonStratifiedStopped & \shevalNA & \shevalNA & \shevalNA & \shevalNA \\
\hline
fresh & \shevalPass & \shevalPass & \shevalPass & \shevalPass & \shevalPass & \shevalPass & \shevalPass & \shevalPass & \shevalPass & \shevalPass & \shevalPass & \shevalPass & \shevalPass & \shevalPass \\
dual1 & \shevalPass & \shevalPass & \shevalPass & \shevalPass & \shevalPass & \shevalCyclesDetectedConformant & \shevalPass & \shevalCyclesDetectedStopped & \shevalPass & \shevalPass & \shevalPass & \shevalFail & \shevalPass & \shevalFail \\
dual2 & \shevalPass & \shevalPass & \shevalPass & \shevalCyclesDetectedConformant & \shevalPass & \shevalCyclesDetectedConformant & \shevalPass & \shevalCyclesDetectedStopped & \shevalPass & \shevalPass & \shevalPass & \shevalFail & \shevalPass & \shevalFail \\
\hline
norec1 & \shevalPass & \shevalPass & \shevalPass & \shevalPass & \shevalPass & \shevalPass & \shevalPass & \shevalPass & \shevalPass & \shevalPass & \shevalPass & \shevalPass & \shevalPass & \shevalPass \\
norec2 & \shevalFail & \shevalFail & \shevalFail & \shevalFail & \shevalFail & \shevalFail & \shevalFail & \shevalFail & \shevalFail & \shevalFail & \shevalFail & \shevalFail & \shevalFail & \shevalFail \\
\hline
\end{tabular}
\caption{
Results of ShEx and SHACL validators for recursive shapes catalogues:\\
\shevalPass{} = all nodes conform to the expected shapes (\lstinline|conforms = true|)\\
\shevalFail{} = some nodes don't conform to their expected shapes (\lstinline|conforms = false|)\\
\shevalTimeout{} = timeout error during validation\\
\shevalCyclesDetectedStopped{} = recursive cycles detected and validation stopped\\
\shevalCyclesDetectedConformant{} = recursive cycles detected, but validator continued and returned (\lstinline|conforms=true|)\\
\shevalCyclesDetectedNonConformant{} = recursive cycles detected, but validator continued and found errors (\lstinline|conforms=false|)\\
\shevalCyclesDetectedCrashed{} = engine detected recursive cycles, attempted the validation and crashed\\
\shevalNonStratifiedStopped{} = engine detected non-stratified recursive shapes and stopped \\
\shevalNA{} = no expected semantics, 
}
\label{tab:sheval-results}
\end{table}

\section{Discussion} \label{sec:discussion}

Using Table~\ref{tab:sheval-results}, we can see which validators are consistent with 
 which semantics. 
 In the case of the non-recursive control tests \texttt{norec1} and \texttt{norec2} 
 all validators behave as expected.

For recursive shapes and non-stratified shape catalogs, we see that the three 
ShEx validators, rudof, Jena ShEx and ShEx-S, are consistent with \GFP, 
as prescribed by the official semantics~\cite{PBGK19}. 

The situation is less uniform in the SHACL ecosystem. 
  pySHACL detects cycles in most of the cases except in \texttt{reach2} 
  but it continues trying to validate and in most of the cases it returns a 
  \lstinline|conforms=true| except in \texttt{example \ref{example_shapes}} and \texttt{bsep3} 
  in which cases it gives an error. 
  SHACL-s fails for example~\ref{example_shapes} and bsep3 and has a timeout error. 
  The semantics seems to follow a \GFP. 
Jena SHACL detects recursive cycles, and continues attempting the validation returning 
  a conforming validation report following the \GFP semantics. 
  The behaviour of SHACL-TQ is a bit different, as it is not consistent 
  with either \GFP, \LFP, bSMS, or cSMS, as can be seen. 
SHACL-TQ thus seems to follow a unique semantics which would require a more in depth 
  look to the algorithm implemented and the source code. 
  Regarding rudof SHACL, it has a flag to indicate whether it stops with recursive shapes (rudof no cycles) 
  and another one to indicate which semantics to use, \GFP (marked as brave) or \LFP (marked as cautious), 
  and the behaviour of rudof SHACL is consistent with the semantics indicated by the flag.

In the case of non-stratified catalogs, 
ShEx-S and rudof (ShEx and SHACL) correctly reject such inputs. 
Jena ShEx does not reject them and attempts to validate, 
answering that the graph validates in some cases or rejects in others. 
\texttt{cons1} asks for a node to be assigned two shapes, where by definition one negates the other. 
A pass in this context means reporting a validation error, as no logically consistent shape 
assignment can satisfy the desired target assignment. 
\texttt{cons2} is a similar test, but involves two nodes instead of just one. 
Maybe this asymmetric behaviour of Jena ShEx for two very similar tests can be 
attributed to a software bug. 
For non-stratified catalogs, except rudof, the rest of the SHACL validators 
attempt the validation and return non-conformance.

In the case of the duality tests, the ShEx validators are consistent with \GFP{} 
 and the SHACL validators also seem to follow \GFP{}.

Our test results do not prove that any of the validators follow a given semantics. 
They just prove that some implementations do not follow a given semantics, 
 and they give us hints as to how these implementations behave. 
Claiming that their semantics follows \GFP{}, \LFP{}, bSMS, or cSMS 
requires knowledge of the algorithm each system implements, 
and a formal study of said algorithm.

\section{Related work} \label{sec:related}

This paper can be considered a sequel of~\cite{recursive_shapes} with a focus on a more practical description of the framework used in that paper for the experiments.  
That paper was preceded by a previous paper focused on finding a common foundation of schema languages like ShEx, SHACL and PG-Schema without recursion\cite{www25}. 
Combining recursive shapes and negation have been a significant topic of recent research. 
This is especially pronounced for SHACL, where the W3C recommendation~\cite{KK17} left the semantics of recursion undefined.  
The challenges of recursion were first formally addressed by Corman et al.~\cite{CRS18,CFRS19}, who proposed a \emph{supported model semantics} based on first-order logic.
Subsequent work has explored various semantics inspired by paradigms from fixed-point logics and 
logic programming~\cite{ACORSS20,OS24,BJ21,AhmetajLOS22,PKM22}. 
In contrast to SHACL, the semantics of recursion in ShEx has been consistently defined via \emph{greatest fixed-points}~\cite{BGP17,SBG15}. 
A previous minimal language common to ShEx and SHACL was the S language defined in~\cite{GGFE19} which was an inspiration for the Simple Shape Language.

There have been several efforts to define benchmarks and test suites for shape languages. One of the earliest was the Webindex benchmark proposed in~\cite{labragayo2017validating} which included both a ShEx and SHACL generator benchmark tool with recursive shapes. 
 Those shapes were inspired by a real-world use case and didn't explore the different semantics associated. 

SHACL recommendation refers to the SHACL test suite~\cite{SHACLTestsuite17} which contains a set of test cases for SHACL engines. 
 The test suite doesn't include tests for recursion, and in case future work on SHACL tackled the need for recursive validation, 
 it could be extended with the recursive test suite presented in this paper. 
The ShEx test suite~\cite{shextest} contains more than one thousand test cases for ShEx engines, including some recursive shapes, 
 but it doesn't explore different semantics of recursion apart of GFP.



\section{Conclusions} \label{sec:conclusions}

We described \textsc{sheval}, a tool to run test suites for the evaluation of different 
 shape engines. 
 It has been used to evaluate and understand the differences in the 
  implementation of recursive shapes in ShEx and SHACL. 
 \textsc{sheval} provides a framework for testing and comparing 
  the behaviour of different shape engines against a set of expectations, 
  helping to identify inconsistencies and potential issues, 
  and providing a basis for further research and development 
  in the field of shape-based validation of RDF data. 
  The tool is easily customizable and generates a detailed report 
   which can be serialized in YAML, CSV or LaTeX so it can be later analysed.  

Future work includes the addition of other ShEx and SHACL implementations. 
 Another line for future work is to add and define additional test suites 
  and cover other aspects of shape-based validation, 
  such as expressiveness, interoperability, performance, explainability or usability.


\paragraph{Acknowledgment}
 This was partially supported by the Province of Bolzano and FWF through project OnTeGra (DOI 10.55776/PIN8884924, Savkovi\'c).

\begin{acknowledgments}
This work has been partially funded by the project SHAKIGRA: Shaping Knowledge and Interoperable Graphs, code: NAC-ES-PUB-ASV-2025 PID2024-157010OBI00, from the Spanish Research Agency, the regional project with code SEK25-GRU-GIC-24-089
and COST Action CA23147 GOBLIN - Global Network on Large-Scale, Cross-domain and Multilingual Open
Knowledge Graphs (Jose E. Labra-Gayo).
This was partially supported by the Province of Bolzano and FWF through project OnTeGra (DOI 10.55776/PIN8884924, Savkovi\'c).
\end{acknowledgments}

\begin{aideclaration}
During the preparation of this work, the author(s) used a combination of ChatGPT, Copilot and Claude Code in order to: Grammar and spelling check, Paraphrase and reword and Improve writing style. After using these tools, the authors reviewed and edited the content as needed and take full responsibility for the publication’s content.
\end{aideclaration}

\bibliography{references}

\end{document}